\documentclass[10pt,
          english,
           aps,
            longbibliography,
            superscriptaddress,
         twocolumn,
           prx]{revtex4-2}

\usepackage[T1]{fontenc}
\usepackage{color}
\usepackage{xcolor}
\usepackage{mathtools}
\usepackage{amsmath,amssymb,mathrsfs}
\usepackage{graphicx}
\graphicspath{{Figures/}}
\usepackage{natbib}
\usepackage{braket}
\usepackage[normalem]{ulem}
\usepackage{float}
\usepackage{dcolumn}
\usepackage{bm}
\usepackage{hyperref}
\usepackage[mathlines]{lineno}
\usepackage{caption}
\usepackage{subcaption}
\usepackage{verbatim}
\begin{document}

\title{Nitrogen-Vacancy Centers in Epitaxial Laterally Overgrown Diamond: Towards Up-scaling of Color Center-based Quantum Technologies}

\author{Nimba Oshnik}
\affiliation{Department of Physics and Research Center OPTIMAS, Rheinland-Pfälzische Technische Universität Kaiserslautern-Landau, 67663 Kaiserslautern, Germany}
\affiliation{Institute for Quantum Control (PGI-8), Forschungszentrum Jülich, DE.}
\affiliation{Institute for Quantum Computation and Analytics (PGI-12), Forschungszentrum Jülich, DE.}

\author{Sebastian Westrich}%
\affiliation{Department of Physics and Research Center OPTIMAS, Rheinland-Pfälzische Technische Universität Kaiserslautern-Landau, 67663 Kaiserslautern, Germany}

\author{Nina Burmeister}%
\affiliation{Department of Physics and Research Center OPTIMAS, Rheinland-Pfälzische Technische Universität Kaiserslautern-Landau, 67663 Kaiserslautern, Germany}
\affiliation{current address: Universität des Saarlandes, Campus Homburg, 66421 Homburg }

\author{Oliver Roman Opaluch}%
\affiliation{Department of Physics and Research Center OPTIMAS, Rheinland-Pfälzische Technische Universität Kaiserslautern-Landau, 67663 Kaiserslautern, Germany}

\author{Lahcene Mehmel}
\affiliation{Department of Physics and Research Center OPTIMAS, Rheinland-Pfälzische Technische Universität Kaiserslautern-Landau, 67663 Kaiserslautern, Germany}
\affiliation{Laboratoire des Sciences des Procédés et des Matériaux – LSPM - Villetaneuse, FR.}

\author{Riadh Issaoui}
\affiliation{Laboratoire des Sciences des Procédés et des Matériaux – LSPM - CNRS, Université Sorbonne Paris Nord, FR.}

\author{Alexandre Tallaire}
\affiliation{Laboratoire des Sciences des Procédés et des Matériaux – LSPM - CNRS, Université Sorbonne Paris Nord, FR.}
\affiliation{Ecole Nationale Supérieure de Chimie de Paris - Chimie Paristech, Université Paris Sciences et Lettres (PSL), FR.}

\author{Ovidiu Brinza}
\affiliation{Laboratoire des Sciences des Procédés et des Matériaux – LSPM - CNRS, Université Sorbonne Paris Nord, FR.}

\author{Jocelyn Achard}
\affiliation{Laboratoire des Sciences des Procédés et des Matériaux – LSPM - CNRS, Université Sorbonne Paris Nord, FR.}

\author{Elke Neu}
\email{nruffing@rptu.de}
\affiliation{Department of Physics and Research Center OPTIMAS, Rheinland-Pfälzische Technische Universität Kaiserslautern-Landau, 67663 Kaiserslautern, Germany}

\begin{abstract}
 Providing high-quality, single-crystal diamond (SCD) with a large area is desirable for up-scaling quantum technology applications that rely on color centers in diamond. Growth methods aiming to increase the area of SCD are an active research area. Native color centers offer a sensitive probe for local crystal quality in such novel materials e.g., via their reaction to stress. In this work, we investigate individual native nitrogen-vacancy (NV) centers in SCD layers manufactured via laterally overgrowing hole arrays in a heteroepitaxially grown large-scale substrate.  Heteroepitaxy has become a common tool for growing large SCDs; however, achieving the high crystal quality needed for quantum applications remains a challenge.  In the overgrown layer, we identify NV centers with spin-decoherence times in the order of hundreds of µs, comparable to high-purity homoepitaxial SCD. We quantify the effective crystal stress in different regions of the overgrown layer, indicating a low stress overall and a stress reduction in the diamond layer above the holes.  
\end{abstract}

\maketitle

\section{\label{sec:1}Introduction}

Enhancing the sensitivity of quantum sensors based on color centers in diamond is an active field of research. Color centers in diamond, especially negatively-charged nitrogen-vacancy (NV) centers, are atomically small sensors for local magnetic fields.\footnote{Unless otherwise stated, NV center refers to the negative charge state throughout the manuscript.} Optimizing their sensitivity allows enhancing their applicability in various fields.  Besides other parameters, it is crucial to obtain NV centers with highly coherent spins and well-defined quantum states.  On the one hand, complex spin manipulation methods can be utilized to improve the performance of diamond based sensors ~\cite{barry2020sensitivity,rembold_introduction_2020}. On the other hand, material advancements contribute to tackling the challenge of protecting the quantum systems from inherent decoherence~\cite{balasubramanian2009ultralong,jahnke2012long,smith2019colour}. However, most work still relies on homoepitaxial diamond. Although homoepitaxial CVD (chemical vapor deposition) growth of diamond is the gold standard for high-quality, high-purity single crystal diamond, the size of these diamonds is typically limited to a few millimeters by the availability of diamond substrates (typically HPHT diamonds). In 2017, the growth of millimeter thick diamond wafers with lateral size up to 10 cm has been demonstrated using heteroepitaxy~\cite{schreck2017ion}. However, our own work on native, individual NV centers in this material revealed a limited coherence times~\cite{nelz2019toward}. Such wafer scale material, however, can serve as a large-area, at least centimeter sized, substrate for additional CVD steps. 

\begin{figure*}[ht!]
     \centering
     \includegraphics[clip, trim=0.0cm 3.1cm 0.0cm 2.1cm,width=0.9\textwidth]{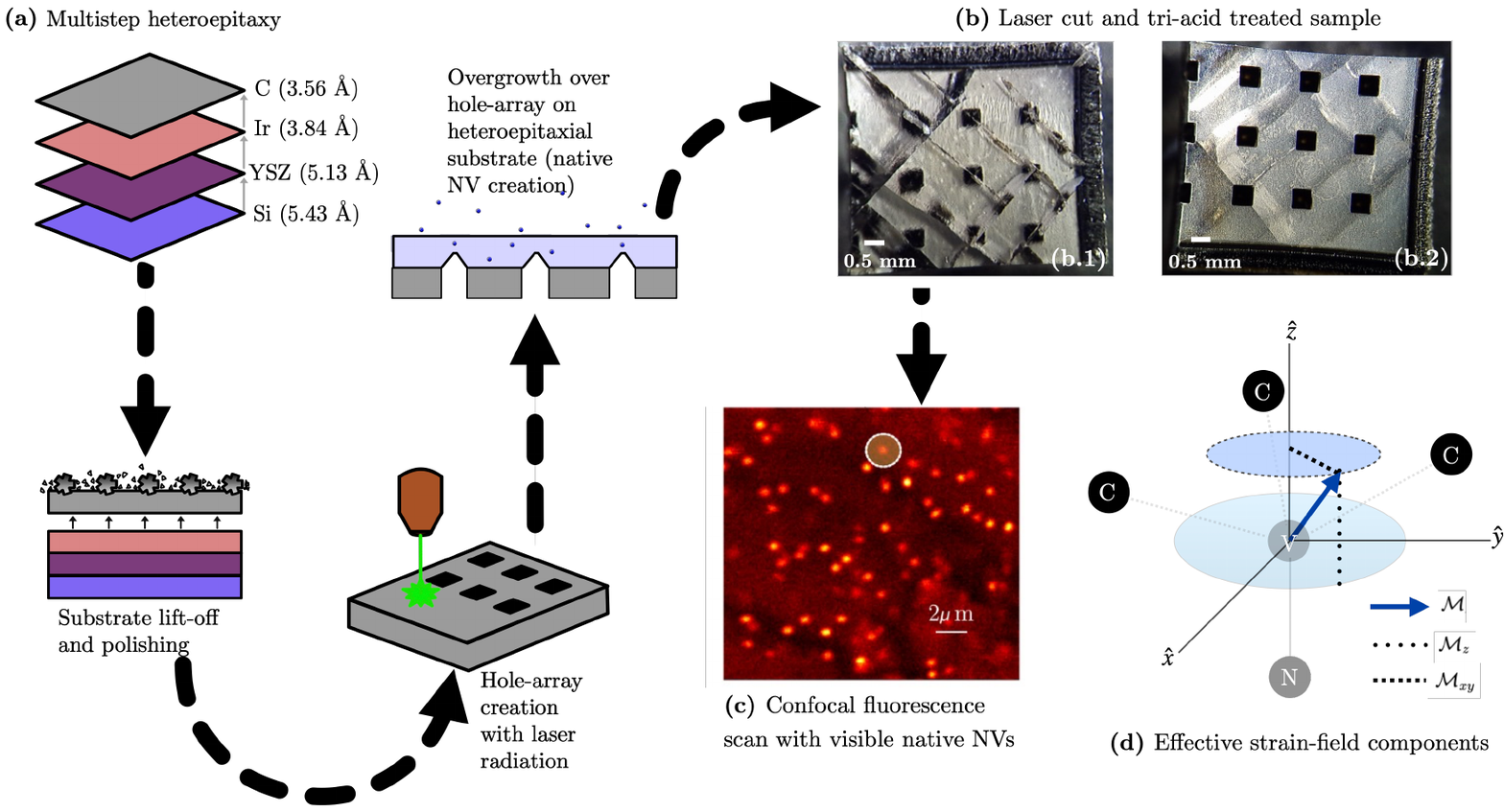}
     \caption{Manufacturing process of the diamond layer investigated in this work:(a) Heteroepitaxy growth process of the commercial substrate; multi layer substrates Ir/YSZ/Si are employed to grow large single crystal diamonds. The multilayer system is chosen as the lattice constant of Ir closely matches the lattice constant of diamond, the YSZ interlayer allows the growth of suitable Ir films on Si (lattice constants given in the figure). Subsequent substrate lift-off and polishing is done to prepare the base crystal for further growth. An array of 500 µm $\times$ 500 µm sized quadratic holes are created using laser cutting. The array field is laterally overgrown using optimized conditions to obtain a $\sim$500 µm thick layer on the 1 cm $\times$ 1 cm substrate. The diamond is laser cut into four parts, one of them is used in this study. (b)  Optical microscope image of the samples used in this study after tri-acid cleaning (see text). Overgrown side (b.1) and substrate side (b.2) of the sample show growth related features. (c) Fluorescence scan of overgrown surface, individual, native NVs are clearly visible as isolated spots (laser excitation at 520 nm). Single NVs are identified and used for study of crystal properties. (d) Single NV coordinate system, showing effective stress $\mathcal{M}$ and its components; a complete description of $\mathcal{M}$ in the 3D space requires measurements with NV oriented along the 4 possible equivalent orientations. Defects like dislocations in the lattice result in a change in the local effective stress.}
     \label{fig:info}
\end{figure*}

Additionally, one has to keep in mind that dislocations, which are present in comparably high density in the wafer scale material, propagate from the substrate to the overgrown layer and induce stress in the overgrown layers. Stress inhomogeneity, in turn, can limit the performance of diamond-based quantum sensors, see for example,~\cite{doherty2012theory}. In this work, we study a CVD diamond layer manufactured via laterally overgrowing hole-arrays in a heteroepitaxially-grown substrate, this approach has been shown to reduce the propagation of dislocations into the overgrown layer in our own previous work ~\cite{mehmel_dislocation_2021}. The approach in reference~\cite{mehmel_dislocation_2021} estimates the dislocation density by observing etch pits forming at the positions of dislocations after plasma exposure of the surface, along with the study of Raman spectra of the diamond sample. Methods like x-ray topography and tunneling electron microscopy can also be used to image dislocations~\cite{schreck2017ion}. Here, we follow a route directly tailored to assess the usability of the diamond for quantum applications via investigation of native NV centers in the diamond layers, as these versatile quantum sensors react to local crystal stress. We here focus on native centers instead of creating color centers via implantation as they are probes for the native, as grown material. In contrast, color center implantation creates additional damage to the diamond and the implanted color centers' properties will partly reflect this damage as well as its repair via annealing. As the dislocation density affects the local stress in the crystal, a corresponding response in the properties of the native NV centers is expected. In this context, we study the variation in NV spin resonance transition frequencies and determine the dephasing and decoherence times ($T_2^\ast$ and $T_2$ respectively) of individual NV centers. In addition to affecting spin-coherence and potentially also hindering nanostructuring processes, higher dislocation densities in the crystal shift the zero-field parameters and can also manifest as low-frequency noise which lowers $T_2^\ast$~\cite{tsuji2024extending}. Thus, low-stress, large-area single crystals are a requirement for diamond-based quantum technologies and nanofabrication. The following section elaborates the sample properties and experimental methods (Sec.~\ref{sec:2}). The findings of the optically detected magnetic resonance (ODMR) experiments,  spin dephasing and decoherence are reported in Sec.~\ref{sec:3}.

\section{\label{sec:2}Sample and Methods}

The sample under consideration is a diamond layer manufactured via laterally overgrowing hole-arrays in a heteroepitaxially-grown substrate. The manufacturing process is sketched in Fig.~\hyperref[fig:info]{1(a)}. The substrate is a $\left<100\right>$ oriented heteroepitaxial diamond substrates (10$\times$10$\times$0.7 mm$^3$; Augsburg Diamond Technology GmbH, Germany). The substrate is grown on an Ir/YSZ/Si template by the provider before being lifted off and polished to optical quality on both sides. An array of quadratic holes with lateral faces oriented along the $\left<100\right>$ direction was then cut through the substrate using laser cutting. This is followed by the growth of a several hundred micrometer-thick CVD diamond layer to fill up the holes. This modified growth method has been developed and tested in the context of growth over diamond substrate in Refs.~\cite{tallaire2017reduction} and~\cite{mehmel_dislocation_2021}. 

In the previous study of similar diamonds, a reduction in the dislocation propagation is reported to obtain diamonds with improved electronic properties. Here, we study native NV centers in the material to enable a crystal stress study, as crystal stress can directly indicate the dislocation density in the vicinity of the defect~\cite{kehayias_imaging_2019}. The original sample is laser-cut into four equivalent pieces, and one of the parts is used for this study.

\begin{figure*}[ht]
     \centering
     \includegraphics[clip, trim=0.0cm 1cm 0.0cm 0.5cm,width=0.9\textwidth]{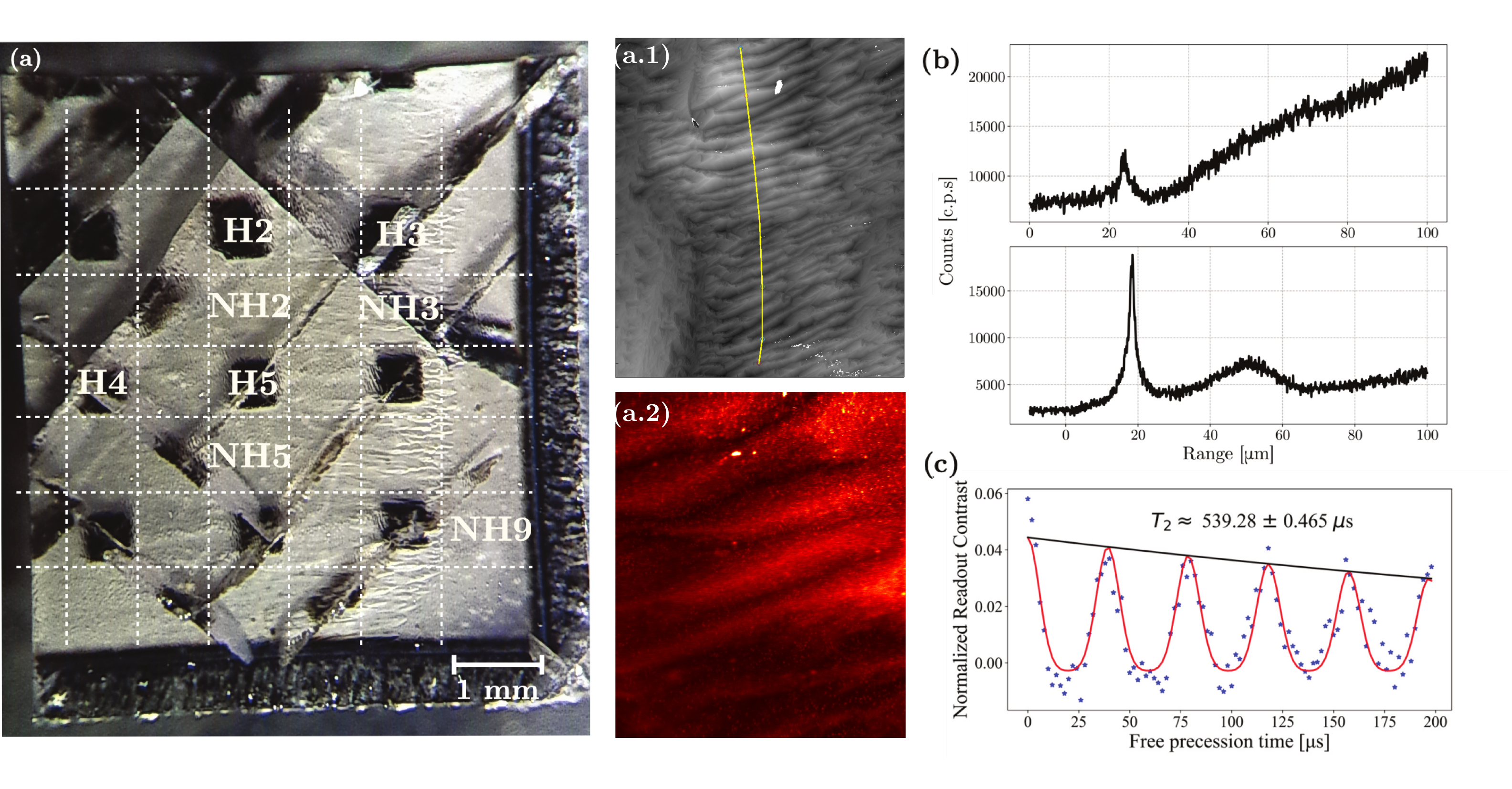}
     \caption{(a) Region demarcation on the overgrown membrane. H indicate laterally overgrown hole regions, whereas regions NH are direct overgrowth over the substrate. (a.1) is surface image obtained with laser scanning confocal microscope. The surface roughness is measured with the line-cut over the growth steps visible in the image, (a.2) shows the growth features as seen under confocal illumination with 520 nm laser light that reveals the single NVs in the membrane. (b) fluorescence vs. axial depth scan before (top) and after (bottom) tri-acid treatment shows reduction of background signal. We attribute the sharp peak observed at around 20 µm to residual back reflected laser light from the sample surface, which is used as a reference for estimating the average depth of the emitters under the surface. Reflected laser light was suppressed by about 5 order of magnitude using a long pass filter. (c) Exemplary analysis of data obtained using the Spin-echo protocol with a NV $\sim$4 µm below the top surface (growth surface) in the region NH9. A T$_2$-time of ca.~540 µs is observed for this particular emitter.}
     \label{fig:region}
\end{figure*}

We use a home-built confocal laser-scanning fluorescence microscope with an excitation laser at 520 nm~\cite{no2022}. The fluorescence is collected through a 600 nm long pass optical filter to suppress back-reflected laser light and fluorescence from the neutral charge state of the NV. Initial confocal fluorescence maps of the overgrown layer revealed significant broad background fluorescence that does not correspond to color centers. We also observed that the measured fluorescence signal was strongly increasing, when we focused deeper into the diamond as visible in Fig.~\hyperref[fig:region]{(b)}. While we could not clearly identify the origin of this fluorescence and its behavior with focus depth, we assume that it arises due to non-diamond carbon and other residues that might be especially pronounced on growth steps or on the backside of the overgrown hole areas. Excitation of such residues by divergent laser light might lead to the observed behavior.  For further investigation, the diamond is treated with a boiling tri-acid mixture (H$_2$SO$_4$ 96 \%, HClO$_4$ 70 \%, HNO$_3$ 65 \%) at 500 $^\circ$C for 1 hour. This treatment removes non-diamond carbon and surface residues, wet-chemically etches non-diamond carbon material, and provides a surface termination with oxygen-containing molecular groups, which stabilizes the negative charge state of the NV centers and improves contrast in ODMR. Fig.~\hyperref[fig:info]{1(b)} shows an optical microscope image of the sample after this cleaning procedure. 

For record and comparison purposes, different labels are assigned to different parts of the sample (Fig.~\hyperref[fig:region]{2(a)}). The regions with keyword H indicate growth with a hole underneath the grown layer, or hole-region in short (H region). NH indicates growth over the substrate, or no-hole region (NH region). The signal enhancement and background reduction is also visible in the vertical z-scan around the sample surface after the tri-acid cleaning (Fig.~\hyperref[fig:region]{2(b)}).

Study of the surface topography of the as-grown layer using a laser scanning confocal microscope (Nanofocus, µsurf explorer, topography resolution 0.1 µm) reveals surface features characteristic for thick CVD-grown diamonds Fig.~\hyperref[fig:region]{2(a.1)}, A mean surface roughness of 1.1$\pm$0.2 µm across the top layer is observed. These growth patterns show an average size of few tens of micrometers. We do not observe differences in this morphology for the H and NH areas. Larger growth steps are observed on the sample (diagonal cut-like structures) even using white light optical microscopy that are further highlighted due to  the stress-relief resulting from laser cutting the sample into four pieces (visible in Fig.~\hyperref[fig:info]{1(b)} and Fig.~\hyperref[fig:region]{2(a)}). We note that the observed surface structure is not detrimental for our confocal fluorescence observations: we do not observe any fluorescence signature from the step-like structure under 520 nm laser excitation (Fig.~\hyperref[fig:info]{1(c)} and~\hyperref[fig:region]{2(a.2)}), indicating that e.g., no non-diamond carbon is present after acid cleaning. 

We thus proceed without polishing the overgrown layer  in order to not change its properties e.g.,\ via inducing additional stress during polishing.
Excitation with 520 nm green laser light reveals  the native NV centers in the layer (Fig.~\hyperref[fig:info]{1(c)}). We identify native single NVs, perform optically detected magnetic resonance (ODMR) experiments, and measurements of spin coherence properties using our home-built confocal microscope ~\cite{no2022}. Random 20 µm $\times$ 20 µm regions are investigated in the H and NH regions, and subsequently random NVs are used for the study to eradicate any bias that may arise from local environment. ODMR experiments are performed without any external bias field to obtain the zero-field parameters of the NV centers. A minimal, unintended stray field is present in the setup because of magnetic parts that have been magnetized during previous experiments and exhibit residual field. The effects of this unintended field on the NV spin transitions might accidentally be interpreted as a stress induced change in the zero-field-parameter. As we cannot fully avoid this field, we need a method to determine the field and to correct for it.   The residual field is determined using a low-stress, commercial, homoepitaxial $\left<100\right>$ diamond (element six, IIa CVD diamond) with an ensemble of NV centers placed next to our sample under investigation, and subsequently accounted for in the studies. In contrast, spin coherence measurements are performed with an external bias field of ca. 8.5 mT along the NV axis. For randomly identified single NVs, the spin-dephasing time $T_2^\ast$ and spin-coherence time $T_2$ are obtained with the Ramsey and Spin-echo protocols, respectively. 

\section{\label{sec:3}Results and Discussion}

\subsection{\label{sec:3.1}Quantifying crystal stress}

One of the primary effects of dislocations in the lattice is a change in the crystal stress field~\cite{lofgren_diamond_2022,korner_influence_2021}. Crystallographic defects such as stacking faults, grain boundaries, and other extended defects lead to similar effects~\cite{blumenau_dislocations_2002,achard2020chemical}. In addition, factors like excess change in pressure, temperature, electric field, along with the effective crystal stress lead to change in the zero-field parameters of the NV center. Hence, in practice, specific experimental design and configuration are needed to distinguish the source of these effects~\cite{dolde2011electric, ivady_pressure_2014}. The experiment involved in this work are performed under ambient conditions, without externally applied electric field. This ensures that any change in the zero-field parameters can be assigned to modified local, effective stress, and thus indicates different local dislocation density. In order to compute the complete stress tensor in the crystal, one can use the information from the four possible NV-axis orientation in the lattice~\cite{trusheim_wide-field_2016, kehayias_imaging_2019}. In the case of single NVs, the knowledge of the precise orientation of the quantization axis with respect to the external magnetic field in the lab frame can provide the same information~\cite{barfuss2019spin}. As the setup and sample restrictions do not allow for such a full computation in this study, only axial and non-axial stress components are quantified for every single NV. In the presence of an external magnetic field and crystal stress, the ground state Hamiltonian of a single NV is given by 
\begin{eqnarray}
\label{NV_ham}
    H =& (D + \mathcal{M}_z)S_z^2 + \gamma_{nv}\vec{B}\cdot\vec{S} \nonumber \\  
    &+\mathcal{M}_x(S_y^2 - S_x^2) + \mathcal{M}_y(S_yS_x + S_xS_y)
\end{eqnarray}
D=2.87 GHz is the zero-field splitting and $\gamma_{nv}$ = 28 GHzT$^{-1}$. $\vec{B}$ is the external magnetic field. $\vec{S} = (S_x, S_y, S_z)$ denote the spin operators, and $\mathcal{M} = (\mathcal{M}_x, \mathcal{M}_y, \mathcal{M}_z)$ indicates the effective electric field associated with crystal-stress. For simplicity, we here set $\hbar=1$. Additional electric fields produced by random or net charge effects are not considered in Eq.~\eqref{NV_ham}. In the limit of $|\vec{B}| \ll D$, the spin-transition frequencies between the $m_s =0$ and $m_s=\pm1$ states obtained via ODMR can be used to compute the effective stress components 
\begin{align}
    \mathcal{M}_{xy} =& ((0.5\times\Delta\nu)^2 - (\gamma_{nv}B_z)^2)^{\frac{1}{2}} \text{ , and} \label{eq:mxy} \\
    \mathcal{M}_z =& 0.5\times(\nu_+ + \nu_-) - D. \label{eq:mz}   
\end{align}

\begin{table*}[ht]
\caption{\label{tab:table1} Average zero-field parameter for NVs in different regions of the overgrown diamond (see Fig.~\hyperref[fig:region]{2(a)}). The values in parentheses give the standard deviation. }
\begin{tabular}{ccccccc}
&\multicolumn{3}{c}{Overgrown substrate (NH-region)}&\multicolumn{3}{c}{Overgrown hole (H-region)}\\
Region-& NH2 & NH3 & NH5 & H2 & H3 & H4 \\ \hline
Avg. D [GHz] & 2.8729 ($\pm$ 0.001) & 2.8716 ($\pm$ 0.002)  & 2.8717 ($\pm$ 0.002) & 2.8705 ($\pm$ 0.001) & 2.8709 ($\pm$ 0.002) & 2.8695 ($\pm$ 0.002) \\
Avg. split [MHz] & 6.85 ($\pm$ 2.81) & 6.98 ($\pm$ 2.53) & 7.27 ($\pm$ 2.59) & 5.55 ($\pm$ 1.17) & 6.53 ($\pm$ 2.54) & 6.43 ($\pm$ 2.61)  \\
$\mathcal \lvert{M}_{z}\rvert$ [MHz]  &2.903&2.138&2.355&0.787&1.623&2.032 \\
$\mathcal{M}_{xy}$ [MHz]  &3.369&3.375&3.581&2.677&2.938&2.798 \\
\end{tabular}
\end{table*}

Where $\mathcal{M}_{xy} = \sqrt{\mathcal{M}_x^2 + \mathcal{M}_y^2}$ is the effective non-axial stress component, $\nu_{+(-)}$ is the $m_s=0 \leftrightarrow 1(-1)$ transition frequency, and $\Delta\nu = \nu_+ - \nu_-$. The spin transition frequencies are obtained via fitting Lorentzian peaks to the ODMR data for individual NV centers in different regions. Around 45 NVs each in the H and NH region in the topmost layer of the overgrown layer are studied in detail. Via observing the laser light reflected at the surface (see Fig.~\hyperref[fig:region]{2(b)}), we ensure to only record the ODMR for NVs less than 5 µm below the growth surface. The large number of observed NVs  ensures random effects are averaged out when looking at collective properties of NVs in a given region. As the dislocation propagation is known to be different in the H and NH region based on previous study~\cite{mehmel_dislocation_2021}, we attribute difference in effective stress to different dislocation density in the regions. Emitters with unknown fluorescence signature, charge instability, and abnormal ODMR spectrum which are occasionally observed are excluded from the study.  The magnitude $\lvert\mathcal{M}\rvert$ and its direction with respect to the NV axis is obtained from axial and non-axial effective stress components. Dislocation cores are typically aligned at the \{111\} plane, and can have directional dependence (partial glide dislocations)~\cite{ghassemizadeh_stability_2022}. However, without knowledge of NV axis orientation, and that the average effect of multiple dislocation cores is observed in the data, it is not possible to quantify the directional dependence of the dislocation from the data. Only the average effects are indicated in the study.

\begin{figure}[h!]
     \centering
     \includegraphics[width=0.45\textwidth]{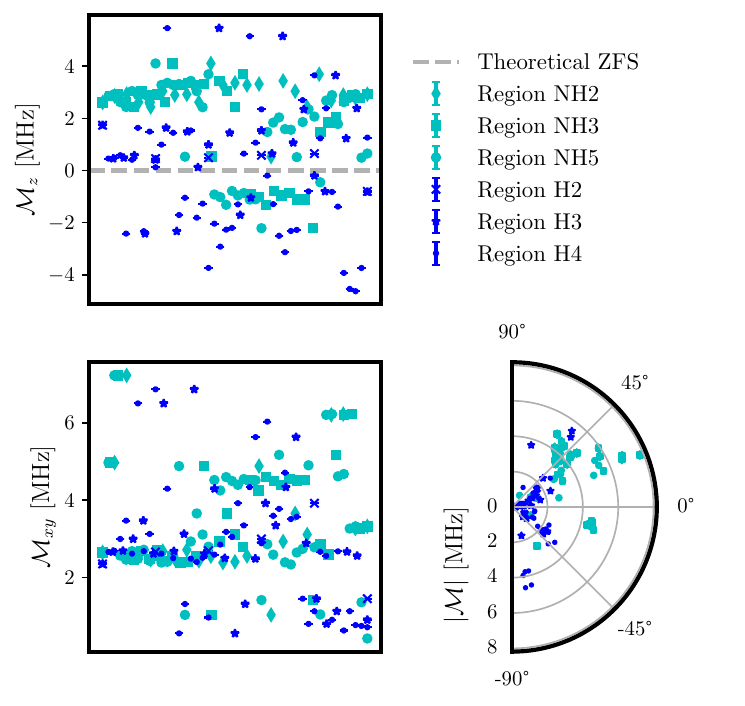}
     \caption{Effective stress components in different regions of study. The zero-field parameters and spin transition frequencies obtained from ODMR measurements are used to compute the $\mathcal{M}$ components for each NV in different regions (see Eq.~\eqref{eq:mxy} and Eq.~\eqref{eq:mz}). The H regions collectively show a variation in the effective stress components for different NVs, implying low and moderate local stress. In comparison, the NVs in the NH regions hint towards a moderate residual stress, where most of the NVs in a given region experience similar effective stress. The magnitude $\lvert\mathcal{M}\rvert$ of the effective stress does not show any pattern for NVs in  the  H regions, with an exception of region H4, where most of the NV show directional effective stress within a 45$^\circ$ to 90$^\circ$ cone with respect to the NV axis. A directional effective stress between -45$^\circ$ to 45$^\circ$ is observed for NVs in the NH regions. See Sec.~\hyperref[sec:3.1]{3A.}, Tab.~\ref{tab:table1}, and Fig.~\hyperref[fig:region]{2(a)} for more details.}
     \label{fig:ODMR-aly}
\end{figure}

Figure~\ref{fig:ODMR-aly} shows the results of the ODMR data analysis: components $M_z$ and $M_{xy}$ indicate minimal effective stress for most of the NVs, which are three orders of magnitudes smaller than NV zero-field parameter $D$. Collectively, large overgrown layer thickness and reduced dislocation propagation limits the stress in the top layer of the diamond. The statistical averages indicate comparatively larger effective stress in the NH regions (Tab.~\ref{tab:table1}), which can also be observed from the bunching of the data points in Fig.~\ref{fig:ODMR-aly}. A direction dependence of the effective stress is observed for the NVs studied in the region NH2, NH3, and NH5. Most of these NVs show $\lvert\mathcal{M}\rvert$ from 3 MHz to 8 MHz, between -45$^\circ$ to 45$^\circ$ cones with respect to the NV axis.

In comparison, there is random variation in the effective stress experienced by the NVs in the H regions, except for region H2, where the effective stress is almost negligible for almost all NVs studied. Region H2 also does not show any directional effective stress, whereas for region H3 and region H4 some NVs indicate a directional effective stress at ca.~45$^\circ$ and ca.~90$^\circ$, respectively. Region H4 also shows higher effective stress in comparison to other H regions. Confocal fluorescence mapping does not contain any signal from the dislocations itself, hence the data inadvertently includes NV in the H region, that are nevertheless very close to a dislocation, or the opposite in the case of the NH region. On average, the observed $\lvert\mathcal{M}_z\rvert$, and $M_{xy}$ values for the NVs in the H region are 20\% to 70\% smaller compared to the NH region (see Tab.~\ref{tab:table1}).

As it is clear from Fig. \ref{fig:ODMR-aly}, individual NV centers experience significantly different stress levels.  To indicate the statistical significance of the above discussed results,  Tab. \ref{tab:table1} lists the standard deviation of the parameters that we extract from the measurement data. For the zero field splitting $D$, we find standard deviations of 1-2 MHz. For the average splitting, the spread between individual NV centers is more significant: the standard deviation is around 30\% of the mean value. We find consistently higher mean values in the NH regions for both the ZFS $D$ and the average splitting between the resonances. To further elaborate statistical significance, we perform a statistical hypothesis test, with the null hypothesis that the mean values for the calculated stress components are the same for NH and H regions. This test results in a T-statistic of ca.~4.8 (P-value=2.5$\times$10$^{-6}$), indicating that the two data sets have significantly different mean values, with the values in the NH region being larger.  From this, we can also conclude that higher average stress components $\vert\mathcal{M}_{xy}\rvert$ and $\lvert\mathcal{M}_z\rvert$ arise in this region with statistical significance.   

\subsection{\label{sec:3.2}Spin coherence properties}

\begin{figure}[h!]
     \centering
     \includegraphics[width=0.35\textwidth]{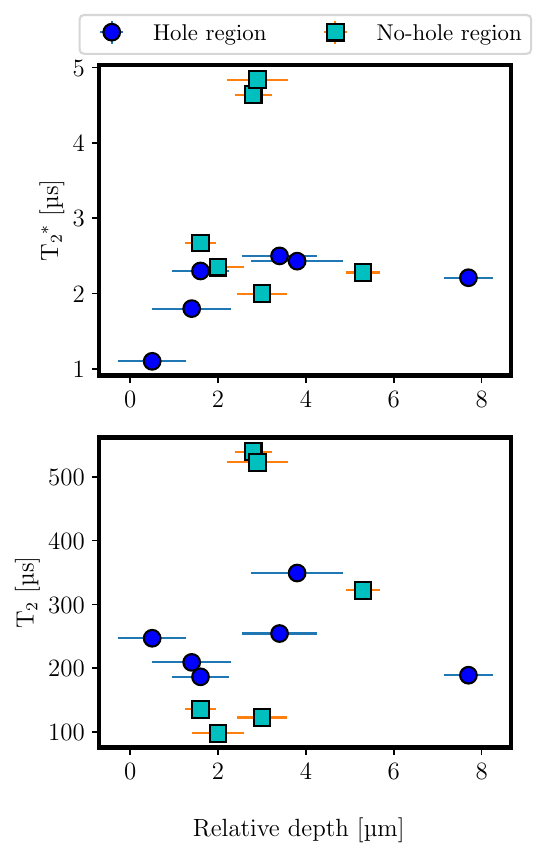}
     \caption{Dephasing (T$_2^{\ast}$) and coherence (T$_2$) times for single NVs in the overgrown layer in H region (Region H5, Fig.~\hyperref[fig:region]{2(a)}) and NH region (Region NH9, Fig.~\hyperref[fig:region]{2(a)}). For each NV, Ramsey and Spin-echo protocols are used to measure T$_2^{\ast}$ and T$_2$, respectively (cf. Tab.~\ref{tab:table2}). Slowly varying noise limits the T$_2^{\ast}$-time, which is enhanced by two orders of magnitude using a refocusing pulse to cancel out the effect of the noise. Near-surface NVs generally have a lower coherence time due to additional noise from surface effects and impurities. The observed characteristic times are on par with high grade homoepitaxial diamonds, however, no observable difference is seen for NVs in the H and NH regions.}
     \label{fig:spind}
\end{figure}

\begin{table*}[ht]
\caption{\label{tab:table2} Measured T$_2^{\ast}$ and T$_2$ times for single NVs in the overgrown layer (Region H5 and NH9 in Fig.~\hyperref[fig:region]{2(a)}). The NVs are ordered by increasing relative depth. The data is fit with suitable modulated exponential functions to obtain the envelope decay times and the errors~\cite{no2022}}. 
\begin{tabular}{ccccc}
&\multicolumn{2}{c}{Overgrown hole (NH5)}&\multicolumn{2}{c}{Overgrown substrate (NH9)} \\
NV label&T$_2^{\ast}$ [µs]&T$_2$ [µs]&T$_2^{\ast}$ [µs]&T$_2$ [µs]\\\hline
NV1 & 1.1 $\pm$ 0.03& 247.01 $\pm$ 0.48 & 2.67 $\pm$ 0.01& 135.48 $\pm$ 0.24\\
NV2 & 1.8 $\pm$ 0.02 & 209.20 $\pm$ 0.34 & 2.35 $\pm$ 0.03 & 97.97 $\pm$ 0.51\\
NV3 & 2.3 $\pm$ 0.04 & 186.60 $\pm$ 0.32 & 4.64 $\pm$ 0.07& 539.28 $\pm$ 0.47\\
NV4 & 2.5 $\pm$ 0.05& 254.42 $\pm$ 0.43 & 4.84 $\pm$ 0.02& 522.87 $\pm$ 0.34\\
NV5 & 2.43 $\pm$ 0.01 & 349.23 $\pm$ 0.5 & 2.00 $\pm$ 0.04& 122.84 $\pm$ 0.87\\
NV6 & 2.21 $\pm$ 0.03& 189.08 $\pm$ 0.71 & 2.28 $\pm$ 0.06& 321.93 $\pm$ 0.90\\ 
\end{tabular}
\end{table*}

Very high dislocation density in the diamond lattice results in high effective electric field, and these defects can also act as charge carrier traps~\cite{schreck2020charge}. This in turn may affect the coherence properties of the NV centers. The effective electric field from crystal-stress, charge traps, and other factors like surface impurities manifest as slowly varying DC noise, that in-homogeneously couples to the NV spin system and leads to faster dephasing, resulting in shorter dephasing time $T_2^\ast$~\cite{tsuji2024extending}. Thus, $T_2^\ast$ is expected to be correlated to the dislocation density or in the case of individual NVs the proximity of the closest dislocation(s). Although several coherent quantum spin manipulation exist that can be applied to decouple the spin from this noise bath and work on the timescale of the spin coherence time $T_2$, it can be crucial, e.g., for quantum sensing methods that aim to detect slowly varying DC fields to have a higher $T_2^\ast$. Following up with the characterization of effective stress component, NVs are studied for their coherence properties in different regions. For this study, region H5 and NH9 are considered. We apply a fixed bias magnetic field of ca. 8.5 mT which is aligned such that it matches one of the 4 equivalent directions of NV center in our diamond sample. Consequently, only NV centers for which this field is aligned  along their quantization axis are used for the measurement. This ensures that the NV spin-transitions are well separated and the level scheme can be approximated as a two-level via adiabatic elimination while minimizing coherence loss induced by spin-mixing due to transverse fields. Ramsey and Spin-echo measurements are performed with single  NV identified in both regions, and the results obtained are shown in Fig.~\ref{fig:spind}. While this approach is optimal to obtain 
$T_2^\ast$ and $T_2$, we cannot localize and re-identify the NV centers we have used under zero-bias field to determine the stress components. With this approach, we thus do not get a full dataset of stress components and coherence times for the same individual NV centers. 
$T_2^\ast$-times in the order of few microseconds are observed for our NVs in the H and NH regions, which are an order of magnitude shorter than in high quality homoepitaxial diamond~\cite{achard2020chemical} and almost two orders of magnitude shorter than that of NVs in isotopically purified crystals~\cite{balasubramanian2009ultralong}. Decoupling the spin from the inhomogeneous noise bath results in $T_2$-times that are typ.~two orders of magnitude larger than the observed $T_2^\ast$-times for the individual NVs. In case of region NH9, $T_2^\ast$-time of ca. 0.5 ms is observed for NVs at a depth of around 2-3 µm. There is no systematic of $T_2^\ast$-times for NVs in the different regions, implying no direct effect of dislocation density on $T_2^\ast$-times. The inhomogeneous coupling of the slow noise contains the collective effect of dislocation density, spin impurities, and other defects. In this context, the findings from the NVs in the two regions indicate uniform inhomogeneous noise. The observed $T_2$-times on the other hand are comparable to high quality homoepitaxial diamond~\cite{achard2020chemical}, greatly superseding our previously reported $T_2$-times for the heteroepitaxial substrate diamonds~\cite{nelz2019toward}. It is noteworthy that a study with ensemble of NVs in similar sample can better quantify the correlation between the dislocation densities occurring at micrometer scale, whereas single NVs are useful to probe local environment at atomic scale. 

Both, ZFS parameters and $T_2^\ast$ times, indicate additional noise in the material as compared to "gold-standard" homoepitaxial material, which is effectively decoupled using echo technique. This may result from a homogeneous dislocation density, where the effective stress is small because contributions from different dislocations cancel. Previous work reports a reduction in the dislocation density, which is reflected in the results. We note that in our previous work, high dislocation densities in the substrate material did not hinder nanofabrication of photonic structures via optimized plasma etching~\cite{nelz2019toward}. 

\section{\label{sec:4}Conclusion}
The findings in this study indicate low effective stress throughout the top layer of our overgrown material,  indicating restrained dislocation propagation from the substrate. The regions overfilling the holes (H regions) show axial (non-axial) stress components ranging between 0.8-2 (2.67-2.94) MHz, in comparison to 2.13-2.90 (3.36-3.58) MHz for regions directly grown on the heteroepitaxial grown diamond as substrate (NH regions), indicating, on average, a 20-70\% reduction in effective crystal stress component. We observe an effective decoherence time (reaching 0.5 ms $T_2$) comparable to high quality homoexpitaxial CVD grown crystal when decoupling the spin defects from slowly varying noise in the crystal. This characteristic is advantageous for the development of robust diamond-based quantum sensors. The presented study is limited in resolving the complete stress-tensor, and thereby the directional effect of the stress due to dislocations. This can be overcome by studying similar sample containing dense ensemble of NVs that are randomly oriented. Despite low stress, further modification of the growth process might be investigated to also enhance $T_2^\ast$ times. 
Our results pave the way towards using large area heteroepitaxial diamonds as growth substrate for up-scaling single crystal diamonds for quantum technologies: For future sensor device fabrication, centimeter-sized heteroepitaxial substrates with an array of millimeter sized, overgrown holes allow for an effective increase in usable diamond area. Despite the fact that only the diamond over the holes will be optimized for sensor fabrication, usable areas will still be significantly larger than standard homoepitaxial substrates limited to few millimeters size. Further processing steps will be necessary to use these diamond for sensor fabrication, e.g. thinning the samples from the backside to retrieve only the topmost layer (closest to the growth surface). While this is a standard technique for small diamond samples, such processes also need to be tested for larger diamonds.   
The effect of polishing and thinning to retrieve only the high quality layer, diamond nanofabrication processes, and the possibility of growing preferentially oriented NV ensemble in the single crystal remains to be tested. 

\section{acknowledgments}

The data used in this manuscript has been made publicly available via the Zenodo repository \url{https://doi.org/10.5281/zenodo.14697917}.

NO acknowledges funding by the Nachwuchsring of RPTU (Research funding 2023, grant no. 82871245). We thank Almax Easylabs for polishing and cutting the diamond. This work was partly funded by the Deutsche Forschungsgemeinschaft  (DFG, German Research Foundation)-- Project No. 429529648--TRR 306 QuCoLiMa (“Quantum Cooperativity of Light and Matter”) and  TRR 173–268565370 ("Spin+X"). EN acknowledges support from the Quantum-Initiative Rhineland-Palatinate (QUIP).

\bibliography{ref}

\begin{thebibliography}{24}%
\makeatletter
\providecommand \@ifxundefined [1]{%
 \@ifx{#1\undefined}
}%
\providecommand \@ifnum [1]{%
 \ifnum #1\expandafter \@firstoftwo
 \else \expandafter \@secondoftwo
 \fi
}%
\providecommand \@ifx [1]{%
 \ifx #1\expandafter \@firstoftwo
 \else \expandafter \@secondoftwo
 \fi
}%
\providecommand \natexlab [1]{#1}%
\providecommand \enquote  [1]{``#1''}%
\providecommand \bibnamefont  [1]{#1}%
\providecommand \bibfnamefont [1]{#1}%
\providecommand \citenamefont [1]{#1}%
\providecommand \href@noop [0]{\@secondoftwo}%
\providecommand \href [0]{\begingroup \@sanitize@url \@href}%
\providecommand \@href[1]{\@@startlink{#1}\@@href}%
\providecommand \@@href[1]{\endgroup#1\@@endlink}%
\providecommand \@sanitize@url [0]{\catcode `\\12\catcode `\$12\catcode `\&12\catcode `\#12\catcode `\^12\catcode `\_12\catcode `\%12\relax}%
\providecommand \@@startlink[1]{}%
\providecommand \@@endlink[0]{}%
\providecommand \url  [0]{\begingroup\@sanitize@url \@url }%
\providecommand \@url [1]{\endgroup\@href {#1}{\urlprefix }}%
\providecommand \urlprefix  [0]{URL }%
\providecommand \Eprint [0]{\href }%
\providecommand \doibase [0]{https://doi.org/}%
\providecommand \selectlanguage [0]{\@gobble}%
\providecommand \bibinfo  [0]{\@secondoftwo}%
\providecommand \bibfield  [0]{\@secondoftwo}%
\providecommand \translation [1]{[#1]}%
\providecommand \BibitemOpen [0]{}%
\providecommand \bibitemStop [0]{}%
\providecommand \bibitemNoStop [0]{.\EOS\space}%
\providecommand \EOS [0]{\spacefactor3000\relax}%
\providecommand \BibitemShut  [1]{\csname bibitem#1\endcsname}%
\let\auto@bib@innerbib\@empty
\bibitem [{Note1()}]{Note1}%
  \BibitemOpen
  \bibinfo {note} {Unless otherwise stated, NV center refers to the negative charge state throughout the manuscript.}\BibitemShut {Stop}%
\bibitem [{\citenamefont {Barry}\ \emph {et~al.}(2020)\citenamefont {Barry}, \citenamefont {Schloss}, \citenamefont {Bauch}, \citenamefont {Turner}, \citenamefont {Hart}, \citenamefont {Pham},\ and\ \citenamefont {Walsworth}}]{barry2020sensitivity}%
  \BibitemOpen
  \bibfield  {author} {\bibinfo {author} {\bibfnamefont {J.~F.}\ \bibnamefont {Barry}}, \bibinfo {author} {\bibfnamefont {J.~M.}\ \bibnamefont {Schloss}}, \bibinfo {author} {\bibfnamefont {E.}~\bibnamefont {Bauch}}, \bibinfo {author} {\bibfnamefont {M.~J.}\ \bibnamefont {Turner}}, \bibinfo {author} {\bibfnamefont {C.~A.}\ \bibnamefont {Hart}}, \bibinfo {author} {\bibfnamefont {L.~M.}\ \bibnamefont {Pham}},\ and\ \bibinfo {author} {\bibfnamefont {R.~L.}\ \bibnamefont {Walsworth}},\ }\href@noop {} {\bibfield  {journal} {\bibinfo  {journal} {Reviews of Modern Physics}\ }\textbf {\bibinfo {volume} {92}},\ \bibinfo {pages} {015004} (\bibinfo {year} {2020})}\BibitemShut {NoStop}%
\bibitem [{\citenamefont {Rembold}\ \emph {et~al.}(2020)\citenamefont {Rembold}, \citenamefont {Oshnik}, \citenamefont {Müller}, \citenamefont {Montangero}, \citenamefont {Calarco},\ and\ \citenamefont {Neu}}]{rembold_introduction_2020}%
  \BibitemOpen
  \bibfield  {author} {\bibinfo {author} {\bibfnamefont {P.}~\bibnamefont {Rembold}}, \bibinfo {author} {\bibfnamefont {N.}~\bibnamefont {Oshnik}}, \bibinfo {author} {\bibfnamefont {M.~M.}\ \bibnamefont {Müller}}, \bibinfo {author} {\bibfnamefont {S.}~\bibnamefont {Montangero}}, \bibinfo {author} {\bibfnamefont {T.}~\bibnamefont {Calarco}},\ and\ \bibinfo {author} {\bibfnamefont {E.}~\bibnamefont {Neu}},\ }\href {https://doi.org/10.1116/5.0006785} {\bibfield  {journal} {\bibinfo  {journal} {AVS Quantum Science}\ }\textbf {\bibinfo {volume} {2}},\ \bibinfo {pages} {024701} (\bibinfo {year} {2020})},\ \bibinfo {note} {publisher: American Vacuum Society}\BibitemShut {NoStop}%
\bibitem [{\citenamefont {Balasubramanian}\ \emph {et~al.}(2009)\citenamefont {Balasubramanian}, \citenamefont {Neumann}, \citenamefont {Twitchen}, \citenamefont {Markham}, \citenamefont {Kolesov}, \citenamefont {Mizuochi}, \citenamefont {Isoya}, \citenamefont {Achard}, \citenamefont {Beck}, \citenamefont {Tissler} \emph {et~al.}}]{balasubramanian2009ultralong}%
  \BibitemOpen
  \bibfield  {author} {\bibinfo {author} {\bibfnamefont {G.}~\bibnamefont {Balasubramanian}}, \bibinfo {author} {\bibfnamefont {P.}~\bibnamefont {Neumann}}, \bibinfo {author} {\bibfnamefont {D.}~\bibnamefont {Twitchen}}, \bibinfo {author} {\bibfnamefont {M.}~\bibnamefont {Markham}}, \bibinfo {author} {\bibfnamefont {R.}~\bibnamefont {Kolesov}}, \bibinfo {author} {\bibfnamefont {N.}~\bibnamefont {Mizuochi}}, \bibinfo {author} {\bibfnamefont {J.}~\bibnamefont {Isoya}}, \bibinfo {author} {\bibfnamefont {J.}~\bibnamefont {Achard}}, \bibinfo {author} {\bibfnamefont {J.}~\bibnamefont {Beck}}, \bibinfo {author} {\bibfnamefont {J.}~\bibnamefont {Tissler}}, \emph {et~al.},\ }\href@noop {} {\bibfield  {journal} {\bibinfo  {journal} {Nature materials}\ }\textbf {\bibinfo {volume} {8}},\ \bibinfo {pages} {383} (\bibinfo {year} {2009})}\BibitemShut {NoStop}%
\bibitem [{\citenamefont {Jahnke}\ \emph {et~al.}(2012)\citenamefont {Jahnke}, \citenamefont {Naydenov}, \citenamefont {Teraji}, \citenamefont {Koizumi},\ and\ \citenamefont {Isoya}}]{jahnke2012long}%
  \BibitemOpen
  \bibfield  {author} {\bibinfo {author} {\bibfnamefont {K.}~\bibnamefont {Jahnke}}, \bibinfo {author} {\bibfnamefont {B.}~\bibnamefont {Naydenov}}, \bibinfo {author} {\bibfnamefont {T.}~\bibnamefont {Teraji}}, \bibinfo {author} {\bibfnamefont {S.}~\bibnamefont {Koizumi}},\ and\ \bibinfo {author} {\bibfnamefont {J.}~\bibnamefont {Isoya}},\ }\href@noop {} {\bibfield  {journal} {\bibinfo  {journal} {Applied physics letters}\ }\textbf {\bibinfo {volume} {101}} (\bibinfo {year} {2012})}\BibitemShut {NoStop}%
\bibitem [{\citenamefont {Smith}\ \emph {et~al.}(2019)\citenamefont {Smith}, \citenamefont {Meynell}, \citenamefont {Bleszynski~Jayich},\ and\ \citenamefont {Meijer}}]{smith2019colour}%
  \BibitemOpen
  \bibfield  {author} {\bibinfo {author} {\bibfnamefont {J.~M.}\ \bibnamefont {Smith}}, \bibinfo {author} {\bibfnamefont {S.~A.}\ \bibnamefont {Meynell}}, \bibinfo {author} {\bibfnamefont {A.~C.}\ \bibnamefont {Bleszynski~Jayich}},\ and\ \bibinfo {author} {\bibfnamefont {J.}~\bibnamefont {Meijer}},\ }\href@noop {} {\bibfield  {journal} {\bibinfo  {journal} {Nanophotonics}\ }\textbf {\bibinfo {volume} {8}},\ \bibinfo {pages} {1889} (\bibinfo {year} {2019})}\BibitemShut {NoStop}%
\bibitem [{\citenamefont {Schreck}\ \emph {et~al.}(2017)\citenamefont {Schreck}, \citenamefont {Gsell}, \citenamefont {Brescia},\ and\ \citenamefont {Fischer}}]{schreck2017ion}%
  \BibitemOpen
  \bibfield  {author} {\bibinfo {author} {\bibfnamefont {M.}~\bibnamefont {Schreck}}, \bibinfo {author} {\bibfnamefont {S.}~\bibnamefont {Gsell}}, \bibinfo {author} {\bibfnamefont {R.}~\bibnamefont {Brescia}},\ and\ \bibinfo {author} {\bibfnamefont {M.}~\bibnamefont {Fischer}},\ }\href@noop {} {\bibfield  {journal} {\bibinfo  {journal} {Scientific reports}\ }\textbf {\bibinfo {volume} {7}},\ \bibinfo {pages} {44462} (\bibinfo {year} {2017})}\BibitemShut {NoStop}%
\bibitem [{\citenamefont {Nelz}\ \emph {et~al.}(2019)\citenamefont {Nelz}, \citenamefont {G{\"o}rlitz}, \citenamefont {Herrmann}, \citenamefont {Slablab}, \citenamefont {Challier}, \citenamefont {Radtke}, \citenamefont {Fischer}, \citenamefont {Gsell}, \citenamefont {Schreck}, \citenamefont {Becher} \emph {et~al.}}]{nelz2019toward}%
  \BibitemOpen
  \bibfield  {author} {\bibinfo {author} {\bibfnamefont {R.}~\bibnamefont {Nelz}}, \bibinfo {author} {\bibfnamefont {J.}~\bibnamefont {G{\"o}rlitz}}, \bibinfo {author} {\bibfnamefont {D.}~\bibnamefont {Herrmann}}, \bibinfo {author} {\bibfnamefont {A.}~\bibnamefont {Slablab}}, \bibinfo {author} {\bibfnamefont {M.}~\bibnamefont {Challier}}, \bibinfo {author} {\bibfnamefont {M.}~\bibnamefont {Radtke}}, \bibinfo {author} {\bibfnamefont {M.}~\bibnamefont {Fischer}}, \bibinfo {author} {\bibfnamefont {S.}~\bibnamefont {Gsell}}, \bibinfo {author} {\bibfnamefont {M.}~\bibnamefont {Schreck}}, \bibinfo {author} {\bibfnamefont {C.}~\bibnamefont {Becher}}, \emph {et~al.},\ }\href@noop {} {\bibfield  {journal} {\bibinfo  {journal} {APL Materials}\ }\textbf {\bibinfo {volume} {7}} (\bibinfo {year} {2019})}\BibitemShut {NoStop}%
\bibitem [{\citenamefont {Doherty}\ \emph {et~al.}(2012)\citenamefont {Doherty}, \citenamefont {Dolde}, \citenamefont {Fedder}, \citenamefont {Jelezko}, \citenamefont {Wrachtrup}, \citenamefont {Manson},\ and\ \citenamefont {Hollenberg}}]{doherty2012theory}%
  \BibitemOpen
  \bibfield  {author} {\bibinfo {author} {\bibfnamefont {M.}~\bibnamefont {Doherty}}, \bibinfo {author} {\bibfnamefont {F.}~\bibnamefont {Dolde}}, \bibinfo {author} {\bibfnamefont {H.}~\bibnamefont {Fedder}}, \bibinfo {author} {\bibfnamefont {F.}~\bibnamefont {Jelezko}}, \bibinfo {author} {\bibfnamefont {J.}~\bibnamefont {Wrachtrup}}, \bibinfo {author} {\bibfnamefont {N.}~\bibnamefont {Manson}},\ and\ \bibinfo {author} {\bibfnamefont {L.}~\bibnamefont {Hollenberg}},\ }\href@noop {} {\bibfield  {journal} {\bibinfo  {journal} {Physical Review B—Condensed Matter and Materials Physics}\ }\textbf {\bibinfo {volume} {85}},\ \bibinfo {pages} {205203} (\bibinfo {year} {2012})}\BibitemShut {NoStop}%
\bibitem [{\citenamefont {Mehmel}\ \emph {et~al.}(2021)\citenamefont {Mehmel}, \citenamefont {Issaoui}, \citenamefont {Brinza}, \citenamefont {Tallaire}, \citenamefont {Mille}, \citenamefont {Delchevalrie}, \citenamefont {Saada}, \citenamefont {Arnault}, \citenamefont {Bénédic},\ and\ \citenamefont {Achard}}]{mehmel_dislocation_2021}%
  \BibitemOpen
  \bibfield  {author} {\bibinfo {author} {\bibfnamefont {L.}~\bibnamefont {Mehmel}}, \bibinfo {author} {\bibfnamefont {R.}~\bibnamefont {Issaoui}}, \bibinfo {author} {\bibfnamefont {O.}~\bibnamefont {Brinza}}, \bibinfo {author} {\bibfnamefont {A.}~\bibnamefont {Tallaire}}, \bibinfo {author} {\bibfnamefont {V.}~\bibnamefont {Mille}}, \bibinfo {author} {\bibfnamefont {J.}~\bibnamefont {Delchevalrie}}, \bibinfo {author} {\bibfnamefont {S.}~\bibnamefont {Saada}}, \bibinfo {author} {\bibfnamefont {J.~C.}\ \bibnamefont {Arnault}}, \bibinfo {author} {\bibfnamefont {F.}~\bibnamefont {Bénédic}},\ and\ \bibinfo {author} {\bibfnamefont {J.}~\bibnamefont {Achard}},\ }\href {https://doi.org/10.1063/5.0033741} {\bibfield  {journal} {\bibinfo  {journal} {Applied Physics Letters}\ }\textbf {\bibinfo {volume} {118}},\ \bibinfo {pages} {061901} (\bibinfo {year} {2021})},\ \bibinfo {note} {publisher: American Institute of Physics}\BibitemShut {NoStop}%
\bibitem [{\citenamefont {Tsuji}\ \emph {et~al.}(2024)\citenamefont {Tsuji}, \citenamefont {Sekiguchi}, \citenamefont {Iwasaki},\ and\ \citenamefont {Hatano}}]{tsuji2024extending}%
  \BibitemOpen
  \bibfield  {author} {\bibinfo {author} {\bibfnamefont {T.}~\bibnamefont {Tsuji}}, \bibinfo {author} {\bibfnamefont {T.}~\bibnamefont {Sekiguchi}}, \bibinfo {author} {\bibfnamefont {T.}~\bibnamefont {Iwasaki}},\ and\ \bibinfo {author} {\bibfnamefont {M.}~\bibnamefont {Hatano}},\ }\href@noop {} {\bibfield  {journal} {\bibinfo  {journal} {Advanced Quantum Technologies}\ }\textbf {\bibinfo {volume} {7}},\ \bibinfo {pages} {2300194} (\bibinfo {year} {2024})}\BibitemShut {NoStop}%
\bibitem [{\citenamefont {Tallaire}\ \emph {et~al.}(2017)\citenamefont {Tallaire}, \citenamefont {Brinza}, \citenamefont {Mille}, \citenamefont {William},\ and\ \citenamefont {Achard}}]{tallaire2017reduction}%
  \BibitemOpen
  \bibfield  {author} {\bibinfo {author} {\bibfnamefont {A.}~\bibnamefont {Tallaire}}, \bibinfo {author} {\bibfnamefont {O.}~\bibnamefont {Brinza}}, \bibinfo {author} {\bibfnamefont {V.}~\bibnamefont {Mille}}, \bibinfo {author} {\bibfnamefont {L.}~\bibnamefont {William}},\ and\ \bibinfo {author} {\bibfnamefont {J.}~\bibnamefont {Achard}},\ }\href@noop {} {\bibfield  {journal} {\bibinfo  {journal} {Advanced Materials (Deerfield Beach, Fla.)}\ }\textbf {\bibinfo {volume} {29}} (\bibinfo {year} {2017})}\BibitemShut {NoStop}%
\bibitem [{\citenamefont {Kehayias}\ \emph {et~al.}(2019)\citenamefont {Kehayias}, \citenamefont {Turner}, \citenamefont {Trubko}, \citenamefont {Schloss}, \citenamefont {Hart}, \citenamefont {Wesson}, \citenamefont {Glenn},\ and\ \citenamefont {Walsworth}}]{kehayias_imaging_2019}%
  \BibitemOpen
  \bibfield  {author} {\bibinfo {author} {\bibfnamefont {P.}~\bibnamefont {Kehayias}}, \bibinfo {author} {\bibfnamefont {M.~J.}\ \bibnamefont {Turner}}, \bibinfo {author} {\bibfnamefont {R.}~\bibnamefont {Trubko}}, \bibinfo {author} {\bibfnamefont {J.~M.}\ \bibnamefont {Schloss}}, \bibinfo {author} {\bibfnamefont {C.~A.}\ \bibnamefont {Hart}}, \bibinfo {author} {\bibfnamefont {M.}~\bibnamefont {Wesson}}, \bibinfo {author} {\bibfnamefont {D.~R.}\ \bibnamefont {Glenn}},\ and\ \bibinfo {author} {\bibfnamefont {R.~L.}\ \bibnamefont {Walsworth}},\ }\href {https://doi.org/10.1103/PhysRevB.100.174103} {\bibfield  {journal} {\bibinfo  {journal} {Physical Review B}\ }\textbf {\bibinfo {volume} {100}},\ \bibinfo {pages} {174103} (\bibinfo {year} {2019})},\ \bibinfo {note} {publisher: American Physical Society}\BibitemShut {NoStop}%
\bibitem [{\citenamefont {Pandey}(2022)}]{no2022}%
  \BibitemOpen
  \bibfield  {author} {\bibinfo {author} {\bibfnamefont {N.~O.}\ \bibnamefont {Pandey}},\ }\emph {\bibinfo {title} {Quantum Optimal Control for Quantum Sensing with Nitrogen-vacancy Centers}},\ \href {http://d-nb.info/1274783305} {Ph.D. thesis},\ \bibinfo  {school} {Technische Universit{\"a}t Kaiserslautern} (\bibinfo {year} {2022})\BibitemShut {NoStop}%
\bibitem [{\citenamefont {Löfgren}\ \emph {et~al.}(2022)\citenamefont {Löfgren}, \citenamefont {Öberg},\ and\ \citenamefont {Larsson}}]{lofgren_diamond_2022}%
  \BibitemOpen
  \bibfield  {author} {\bibinfo {author} {\bibfnamefont {R.}~\bibnamefont {Löfgren}}, \bibinfo {author} {\bibfnamefont {S.}~\bibnamefont {Öberg}},\ and\ \bibinfo {author} {\bibfnamefont {J.~A.}\ \bibnamefont {Larsson}},\ }\href {https://doi.org/10.1063/5.0080096} {\bibfield  {journal} {\bibinfo  {journal} {AIP Advances}\ }\textbf {\bibinfo {volume} {12}},\ \bibinfo {pages} {035009} (\bibinfo {year} {2022})}\BibitemShut {NoStop}%
\bibitem [{\citenamefont {Körner}\ \emph {et~al.}(2021)\citenamefont {Körner}, \citenamefont {Urban},\ and\ \citenamefont {Elsässer}}]{korner_influence_2021}%
  \BibitemOpen
  \bibfield  {author} {\bibinfo {author} {\bibfnamefont {W.}~\bibnamefont {Körner}}, \bibinfo {author} {\bibfnamefont {D.~F.}\ \bibnamefont {Urban}},\ and\ \bibinfo {author} {\bibfnamefont {C.}~\bibnamefont {Elsässer}},\ }\href {https://doi.org/10.1103/PhysRevB.103.085305} {\bibfield  {journal} {\bibinfo  {journal} {Physical Review B}\ }\textbf {\bibinfo {volume} {103}},\ \bibinfo {pages} {085305} (\bibinfo {year} {2021})},\ \bibinfo {note} {publisher: American Physical Society}\BibitemShut {NoStop}%
\bibitem [{\citenamefont {Blumenau}\ \emph {et~al.}(2002)\citenamefont {Blumenau}, \citenamefont {Heggie}, \citenamefont {Fall}, \citenamefont {Jones},\ and\ \citenamefont {Frauenheim}}]{blumenau_dislocations_2002}%
  \BibitemOpen
  \bibfield  {author} {\bibinfo {author} {\bibfnamefont {A.~T.}\ \bibnamefont {Blumenau}}, \bibinfo {author} {\bibfnamefont {M.~I.}\ \bibnamefont {Heggie}}, \bibinfo {author} {\bibfnamefont {C.~J.}\ \bibnamefont {Fall}}, \bibinfo {author} {\bibfnamefont {R.}~\bibnamefont {Jones}},\ and\ \bibinfo {author} {\bibfnamefont {T.}~\bibnamefont {Frauenheim}},\ }\href {https://doi.org/10.1103/PhysRevB.65.205205} {\bibfield  {journal} {\bibinfo  {journal} {Physical Review B}\ }\textbf {\bibinfo {volume} {65}},\ \bibinfo {pages} {205205} (\bibinfo {year} {2002})},\ \bibinfo {note} {publisher: American Physical Society}\BibitemShut {NoStop}%
\bibitem [{\citenamefont {Achard}\ \emph {et~al.}(2020)\citenamefont {Achard}, \citenamefont {Jacques},\ and\ \citenamefont {Tallaire}}]{achard2020chemical}%
  \BibitemOpen
  \bibfield  {author} {\bibinfo {author} {\bibfnamefont {J.}~\bibnamefont {Achard}}, \bibinfo {author} {\bibfnamefont {V.}~\bibnamefont {Jacques}},\ and\ \bibinfo {author} {\bibfnamefont {A.}~\bibnamefont {Tallaire}},\ }\href@noop {} {\bibfield  {journal} {\bibinfo  {journal} {Journal of Physics D: Applied Physics}\ }\textbf {\bibinfo {volume} {53}},\ \bibinfo {pages} {313001} (\bibinfo {year} {2020})}\BibitemShut {NoStop}%
\bibitem [{\citenamefont {Dolde}\ \emph {et~al.}(2011)\citenamefont {Dolde}, \citenamefont {Fedder}, \citenamefont {Doherty}, \citenamefont {N{\"o}bauer}, \citenamefont {Rempp}, \citenamefont {Balasubramanian}, \citenamefont {Wolf}, \citenamefont {Reinhard}, \citenamefont {Hollenberg}, \citenamefont {Jelezko} \emph {et~al.}}]{dolde2011electric}%
  \BibitemOpen
  \bibfield  {author} {\bibinfo {author} {\bibfnamefont {F.}~\bibnamefont {Dolde}}, \bibinfo {author} {\bibfnamefont {H.}~\bibnamefont {Fedder}}, \bibinfo {author} {\bibfnamefont {M.~W.}\ \bibnamefont {Doherty}}, \bibinfo {author} {\bibfnamefont {T.}~\bibnamefont {N{\"o}bauer}}, \bibinfo {author} {\bibfnamefont {F.}~\bibnamefont {Rempp}}, \bibinfo {author} {\bibfnamefont {G.}~\bibnamefont {Balasubramanian}}, \bibinfo {author} {\bibfnamefont {T.}~\bibnamefont {Wolf}}, \bibinfo {author} {\bibfnamefont {F.}~\bibnamefont {Reinhard}}, \bibinfo {author} {\bibfnamefont {L.~C.}\ \bibnamefont {Hollenberg}}, \bibinfo {author} {\bibfnamefont {F.}~\bibnamefont {Jelezko}}, \emph {et~al.},\ }\href@noop {} {\bibfield  {journal} {\bibinfo  {journal} {Nature Physics}\ }\textbf {\bibinfo {volume} {7}},\ \bibinfo {pages} {459} (\bibinfo {year} {2011})}\BibitemShut {NoStop}%
\bibitem [{\citenamefont {Ivády}\ \emph {et~al.}(2014)\citenamefont {Ivády}, \citenamefont {Simon}, \citenamefont {Maze}, \citenamefont {Abrikosov},\ and\ \citenamefont {Gali}}]{ivady_pressure_2014}%
  \BibitemOpen
  \bibfield  {author} {\bibinfo {author} {\bibfnamefont {V.}~\bibnamefont {Ivády}}, \bibinfo {author} {\bibfnamefont {T.}~\bibnamefont {Simon}}, \bibinfo {author} {\bibfnamefont {J.~R.}\ \bibnamefont {Maze}}, \bibinfo {author} {\bibfnamefont {I.~A.}\ \bibnamefont {Abrikosov}},\ and\ \bibinfo {author} {\bibfnamefont {A.}~\bibnamefont {Gali}},\ }\href {https://doi.org/10.1103/PhysRevB.90.235205} {\bibfield  {journal} {\bibinfo  {journal} {Physical Review B}\ }\textbf {\bibinfo {volume} {90}},\ \bibinfo {pages} {235205} (\bibinfo {year} {2014})},\ \bibinfo {note} {publisher: American Physical Society}\BibitemShut {NoStop}%
\bibitem [{\citenamefont {Trusheim}\ and\ \citenamefont {Englund}(2016)}]{trusheim_wide-field_2016}%
  \BibitemOpen
  \bibfield  {author} {\bibinfo {author} {\bibfnamefont {M.~E.}\ \bibnamefont {Trusheim}}\ and\ \bibinfo {author} {\bibfnamefont {D.}~\bibnamefont {Englund}},\ }\href {https://doi.org/10.1088/1367-2630/aa5040} {\bibfield  {journal} {\bibinfo  {journal} {New Journal of Physics}\ }\textbf {\bibinfo {volume} {18}},\ \bibinfo {pages} {123023} (\bibinfo {year} {2016})},\ \bibinfo {note} {publisher: IOP Publishing}\BibitemShut {NoStop}%
\bibitem [{\citenamefont {Barfuss}\ \emph {et~al.}(2019)\citenamefont {Barfuss}, \citenamefont {Kasperczyk}, \citenamefont {Kölbl},\ and\ \citenamefont {Maletinsky}}]{barfuss2019spin}%
  \BibitemOpen
  \bibfield  {author} {\bibinfo {author} {\bibfnamefont {A.}~\bibnamefont {Barfuss}}, \bibinfo {author} {\bibfnamefont {M.}~\bibnamefont {Kasperczyk}}, \bibinfo {author} {\bibfnamefont {J.}~\bibnamefont {Kölbl}},\ and\ \bibinfo {author} {\bibfnamefont {P.}~\bibnamefont {Maletinsky}},\ }\href {https://doi.org/10.1103/PhysRevB.99.174102} {\bibfield  {journal} {\bibinfo  {journal} {Physical Review B}\ }\textbf {\bibinfo {volume} {99}},\ \bibinfo {pages} {174102} (\bibinfo {year} {2019})}\BibitemShut {NoStop}%
\bibitem [{\citenamefont {Ghassemizadeh}\ \emph {et~al.}(2022)\citenamefont {Ghassemizadeh}, \citenamefont {Körner}, \citenamefont {Urban},\ and\ \citenamefont {Elsässer}}]{ghassemizadeh_stability_2022}%
  \BibitemOpen
  \bibfield  {author} {\bibinfo {author} {\bibfnamefont {R.}~\bibnamefont {Ghassemizadeh}}, \bibinfo {author} {\bibfnamefont {W.}~\bibnamefont {Körner}}, \bibinfo {author} {\bibfnamefont {D.~F.}\ \bibnamefont {Urban}},\ and\ \bibinfo {author} {\bibfnamefont {C.}~\bibnamefont {Elsässer}},\ }\href {https://doi.org/10.1103/PhysRevB.106.174111} {\bibfield  {journal} {\bibinfo  {journal} {Physical Review B}\ }\textbf {\bibinfo {volume} {106}},\ \bibinfo {pages} {174111} (\bibinfo {year} {2022})},\ \bibinfo {note} {publisher: American Physical Society}\BibitemShut {NoStop}%
\bibitem [{\citenamefont {Schreck}\ \emph {et~al.}(2020)\citenamefont {Schreck}, \citenamefont {{\v{S}}{\v{c}}ajev}, \citenamefont {Tr{\"a}ger}, \citenamefont {Mayr}, \citenamefont {Gr{\"u}nwald}, \citenamefont {Fischer},\ and\ \citenamefont {Gsell}}]{schreck2020charge}%
  \BibitemOpen
  \bibfield  {author} {\bibinfo {author} {\bibfnamefont {M.}~\bibnamefont {Schreck}}, \bibinfo {author} {\bibfnamefont {P.}~\bibnamefont {{\v{S}}{\v{c}}ajev}}, \bibinfo {author} {\bibfnamefont {M.}~\bibnamefont {Tr{\"a}ger}}, \bibinfo {author} {\bibfnamefont {M.}~\bibnamefont {Mayr}}, \bibinfo {author} {\bibfnamefont {T.}~\bibnamefont {Gr{\"u}nwald}}, \bibinfo {author} {\bibfnamefont {M.}~\bibnamefont {Fischer}},\ and\ \bibinfo {author} {\bibfnamefont {S.}~\bibnamefont {Gsell}},\ }\href@noop {} {\bibfield  {journal} {\bibinfo  {journal} {Journal of Applied Physics}\ }\textbf {\bibinfo {volume} {127}} (\bibinfo {year} {2020})}\BibitemShut {NoStop}%
\end{thebibliography}%

\end{document}